\documentclass[english,twocolumn,showpacs,amsmath,amssymb,aps]{revtex4-1}
\usepackage[T1]{fontenc}
\usepackage[latin9]{inputenc}
\usepackage{color}
\usepackage{array}
\usepackage{multirow}
\usepackage{amssymb}
\usepackage{graphicx}

\makeatletter

\providecommand{\tabularnewline}{\\}

 \@ifundefined{textcolor}{}
 {%
   \definecolor{BLACK}{gray}{0}
   \definecolor{WHITE}{gray}{1}
   \definecolor{RED}{rgb}{1,0,0}
   \definecolor{GREEN}{rgb}{0,1,0}
   \definecolor{BLUE}{rgb}{0,0,1}
   \definecolor{CYAN}{cmyk}{1,0,0,0}
   \definecolor{MAGENTA}{cmyk}{0,1,0,0}
   \definecolor{YELLOW}{cmyk}{0,0,1,0}
 }

\makeatother

\usepackage{babel}
\begin{document}

\title{Global empirical potentials from purely rotational measurements}

\author{Nikesh S. Dattani,$^{1,2*}$ L. N. Zack,$^{3}$ Ming Sun,$^{4}$
Erin R. Johnson,$^{5}$ Robert J. Le$\,$Roy,$^{6}$ L. M. Ziurys$^{7}$ }

\affiliation{$^{1}$Physical and Theoretical Chemistry Laboratory, Department
of Chemistry, Oxford University, OX1 3QZ, Oxford, UK, }

\affiliation{$^{2}$Quantum Chemistry Laboratory, Department of Chemistry, Kyoto
University, 606-8502, Kyoto, Japan,}

\affiliation{$^{3}$Department of Chemistry, Wayne State University, 48202, Detroit,
Michigan, USA, }

\affiliation{$^{4}$College of Electronic and Optoelectronic Technology, Nanjing
University of Science and Technology, 210094, Nanjing, China,}

\affiliation{$^{5}$Chemistry and Chemical Biology, School of Natural Sciences,
University of California, Merced, 95343, Merced, California, USA,}

\affiliation{$^{6}$Guelph-Waterloo Centre for Graduate Work in Chemistry and
Biochemistry, University of Waterloo, N2L 3G1, Waterloo, Ontario,
Canada,}

\affiliation{$^{7}$Department of Chemistry, University of Arizona, 85721, Tucson,
Arizona, USA.}
\begin{abstract}
The recent advent of chirped-pulse FTMW technology has created a plethora
of pure rotational spectra for molecules for which no vibrational
information is known. The growing number of such spectra demands a
way to build empirical potential energy surfaces for molecules, without
relying on any vibrational measurements. Using ZnO as an example,
we demonstrate a powerful technique for efficiently accomplishing
this. We first measure eight new ultra-high precision ($\pm2$ kHz)
pure rotational transitions in the $X$-state of ZnO. Combining them
with previous high-precision ($\pm50$ kHz) pure rotational measurements
of different transitions in the same system, we have data that spans
the bottom 10\% of the well. Despite not using any vibrational information,
our empirical potentials are able to determine the size of the vibrational
spacings and bond lengths, with precisions that are more than three
and two orders of magnitude greater, respectively, than the most precise
empirical values previously known, and the most accurate \emph{ab
initio }calculations in today's reach. By calculating the $C_{6},$
$C_{8},$ and $C_{10}$ long-range constants and using them to anchor
the top of the well, our potential is \emph{globally} in excellent
agreement with \emph{ab initio }calculations, without the need for
vibrational spectra and without the need for \emph{any }data in the
top 90\% of the well.
\end{abstract}
\maketitle
Measuring and assigning microwave spectra is now easier than ever.
Plenty of systems are having high-resolution spectra recorded for
the first time, and in many of these systems it is only the pure rotational
transitions that are available (\cite{Perez2013}, \cite{Womack2014},
\cite{Frohman2011}, and \cite{Zack2009b}, just to name a few).
Outstanding examples of recent studies for which measurements of pure
rotational spectra have been made include  PbCl and PbF which are
at sharp focus in the investigation of the electron electric dipole
moment for ruling out alternatives to the Standard Model \cite{Baron2014,Hudson2011},\cite{Norman2014};
and the open-shell diatomic SnI whose spectrum was previously considered
too challenging to study \cite{Zaleski2014}. Furthermore, chirped-pulse
FTMW technology now promises the emergence of yet another wave of
new pure rotational spectra, demanding techniques for extracting the
most information from experiments in the absence of vibrational information. 

Among the long list of systems for which only pure rotational data
is available, one example that has wide-ranging applications across
physics is ZnO. Condensed- and gas-phase studies of zinc oxide based
materials have been under intense focus for several decades \cite{Ozgur2005,Gong2009,*Oymak2012}.
ZnO has a wide band gap ($\sim$ 3.4 eV) and a large exciton binding
energy ($\sim$ 60 meV) \cite{Ozgur2005}, making it an attractive
material for many industrial applications. Bulk ZnO is often used
as a semiconductor and can be doped with other materials to enhance
functionality. Many of these properties make ZnO a more advantageous
choice than GaN, for example, in components for optoelectronics,
electronic circuits \textcolor{red}{}, and spintronics . High electromechanical
coupling constants and robustness also make ZnO-based materials a
popular choice for a large variety of nanostructures. ZnO has become
the subject of an enormous number of experimental and theoretical
studies, and has been surveyed extensively in several recent reviews
\cite{Ozgur2005,Gong2009,*Oymak2012}.

However, despite this extensive amount of work on ZnO, there has been
some hesitation to experimentally study the ZnO monomer. Diatomic
oxides of 3$d$ transition metals are very important for astrophysics
and high-temperature chemistry and consequently have received plenty
of attention from theoreticians and experimentalists \cite{Merer1989,*Barnbaum1996,*Tylenda2005}.
Contrarily, Zinc has often been neglected from such laboratory studies
on $d$-block metal oxides because it has completely filled subshells
and thus is not a true transition metal \cite{Jensen2003}. Additionally,
the metal\textquoteright{}s low electron affinity and high ionization
potential seem to indicate that it would not be reactive with other
atoms, molecules, or ligands. Spectroscopic studies of the ground
$X(1^{1}\Sigma^{+})$ electronic state of monomeric ZnO thus far are
limited to merely one high-resolution study \cite{Zack2009}, where
only rotational information for was obtained, and even this was only
for $J\ge8$ and for just the lowest 5 out of scores of vibrational
states. With over 90\% of the potential completely unexplored, and
a sheer lack of vibrational information, the ambition to build a global
empirical potential all the way up to dissociation would seem arduous
if at all possible, and furthermore, the data in \cite{Zack2009}
started from a rotational level of $J=8$, ruling out the possibility
to obtain even a rotationless potential. 

\begin{figure*}
\caption{\label{fig:potential}\textcolor{black}{\scriptsize Comparison of
our empirical potential with }\textcolor{black}{\emph{\scriptsize ab
initio}}\textcolor{black}{\scriptsize{} points based on Ref. \cite{Sakellaris2010},
and with the long-range theory based on the $C_{m}$ coefficients
calculated in this work and damping functions $D_{m}(r)$ as defined
in Ref. \cite{LeRoy2011}. The global potential is accurate over a
broad range of internuclear distances, despite the data only covering
$\sim$ 10\% of the well, and being only pure rotational transitions.}}

\includegraphics[width=1\textwidth]{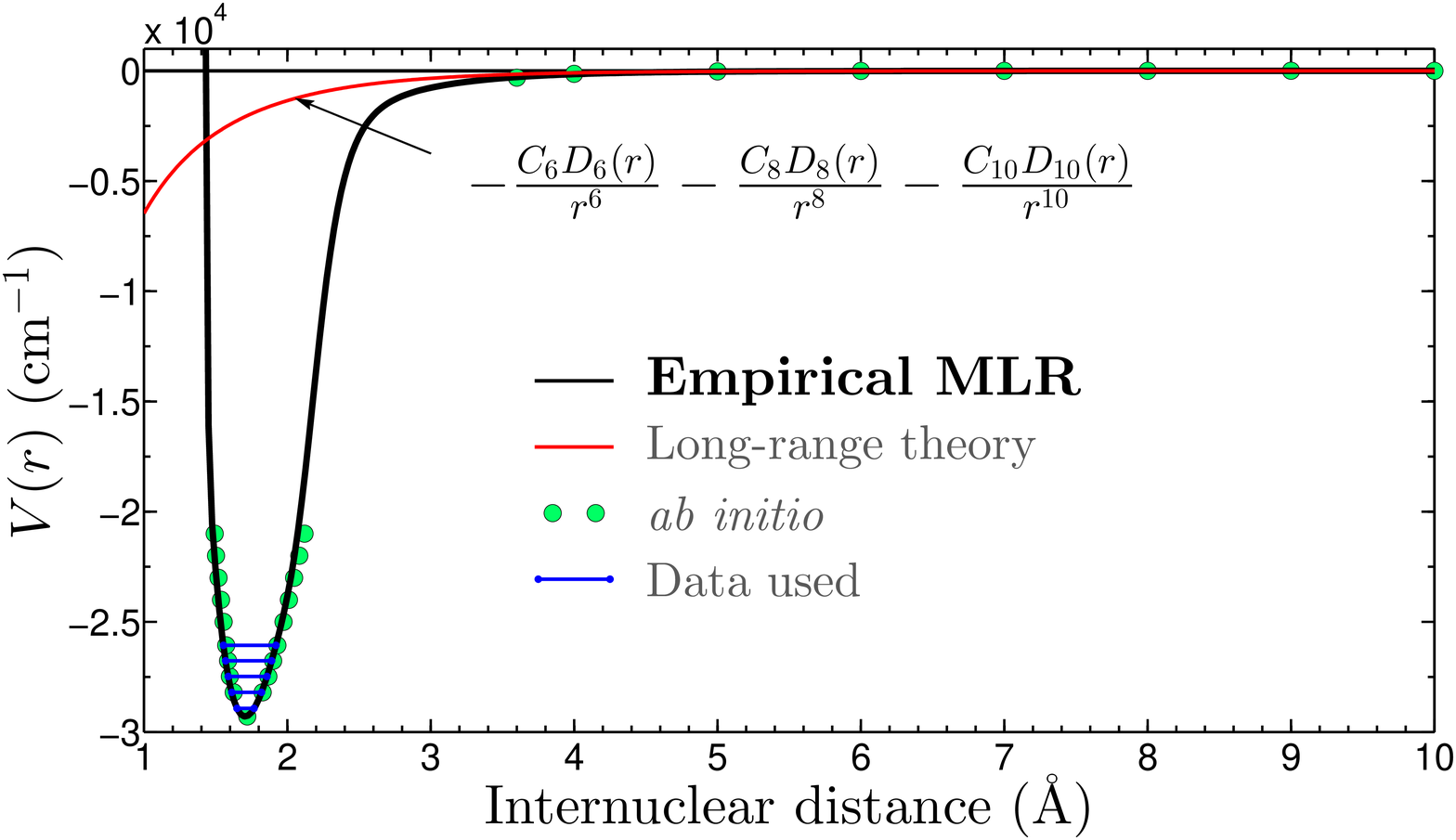}
\end{figure*}

\begin{table}
\caption{\label{tab:newData}\textcolor{black}{\scriptsize Summary of the new
transition energies measured in this work (in MHz), and their uncertainties.}}

\begin{tabular*}{1\columnwidth}{@{\extracolsep{\fill}}cccccccc}
\noalign{\vskip1.5mm}
$v$ & $J^{\prime}$ & $J^{\prime\prime}$ & $^{64}\mbox{Zn}^{16}\mbox{O}$ & $^{66}\mbox{Zn}^{16}\mbox{O}$ & $^{68}\mbox{Zn}^{16}\mbox{O}$ & $^{70}\mbox{Zn}^{16}\mbox{O}$ & Unc.\tabularnewline
\hline 
\noalign{\vskip1mm}
$0$ & $1$ & $0$ & $27072.788$  & $26909.031$  & $26754.756$  & $26609.181$  &  $\pm2$ kHz\tabularnewline
$1$ & $1$ & $0$ & $26843.893$  & $26682.213$  & $-$ & $-$ &  $\pm2$ kHz\tabularnewline
$2$ & $1$ & $0$ & $26615.559$  & $-$ & $-$ & $-$ &  $\pm2$ kHz\tabularnewline
$3$ & $1$ & $0$ & $26387.293$  & $-$ & $-$ & $-$ &  $\pm2$ kHz\tabularnewline
\end{tabular*}

\vspace{-0.6cm}
\end{table}

In this Letter we report new measurements of the $J(1\rightarrow0)$
transitions for isotopologues of the ZnO $X-$state covering all stable
isotopes of Zn with nuclear spin 0. Our new data span various vibrational
states in the range $v=0-3$, and our measurements are more than 
an order of magnitude more precise than the measurements of the much
higher rotational levels reported in \cite{Zack2009}. Early attempts
to build an accurate global potential from purely rotational data
go back to as early as 1995 when Grabow \emph{et al.} fitted various
potential function models to microwave sptectra of NeAr \cite{Grabow1995}.
However, the only potential function models available at the time
were very primitive and different models gave very different potentials.
In 2006, and again in 2008, Grabow and co-workers demonstrated improved
pursuits in this endeavor in a series of excellent papers on on Se-
and Te-based diatomics \cite{Banser2006,*Giuliano2008}, and while
the models used there were more sophisticated, those potentials do
not extrapolate well beyond the data region since these Morse-like
models do not have the correct long-range behavior built in. In 2010
one of us (RJL) compared the well depths predicted by global potentials
based only on microwave data for $v=0$ to those using all data available,
but the full potentials and other predicted properties such as the
vibrational spacing were not compared at that time \cite{Piticco2010}. 

With the new measurements in this work, and a clever choice of a potential
function model that intrinsically promises to yield the correct behavior
in much of the region where data is absent, we are able to build analytic
potentials which are robust globally from the bottom of the well to
the dissociation limit. For each isotopologue, our potentials reproduce
all experimental transition data with an overall standard deviation
well below the experimental uncertainties, and also match well with
state of the art \emph{ab initio} calculations from a recent comprehensive
study \cite{Sakellaris2010}. Our potentials allow us to make various
predictions that were unattainable previously. 

\begin{table*}
\caption{\label{tab:fullDataSet}\textcolor{black}{\scriptsize Summary of the
full data set used in building our empirical potential. Numbers in
columns 3 to 7 indicate the number of data for particular states.}}

\begin{tabular*}{1\textwidth}{@{\extracolsep{\fill}}ccccccccc}
\noalign{\vskip1.5mm}
$J^{\prime\prime}=J^{\prime}-1$ & $v$ & $^{64}\mbox{Zn}^{16}\mbox{O}$ & $^{66}\mbox{Zn}^{16}\mbox{O}$ & $^{67}\mbox{Zn}^{16}\mbox{O}$ & $^{68}\mbox{Zn}^{16}\mbox{O}$ & $^{70}\mbox{Zn}^{16}\mbox{O}$ & Uncertainty & Reference\tabularnewline
\hline 
$0$ & $0$ & $1$ & 1 & - & 1 & 1 &  $\pm2$ kHz & This work\tabularnewline
 & $1$ & $1$ & 1 & - & - & - &  $\pm2$ kHz & This work\tabularnewline
 & $2$ & $1$ & - & - & - & - &  $\pm2$ kHz & This work\tabularnewline
 & $3$ & $1$ & - & - & - & - &  $\pm2$ kHz & This work\tabularnewline
$8-9$ & $0$ & $2$ & 2 & 2 & 2 & 2 & $\pm50$ kHz & \cite{Zack2009}\tabularnewline
$13$ & $0$ & 1 & 1 & 1 & 1 & 1 & $\pm50$ kHz & \cite{Zack2009}\tabularnewline
 & $1$ & 1 & 1 & 1 & 1 & - & $\pm50$ kHz & \cite{Zack2009}\tabularnewline
 & $2$ & 1 & 1 & - & - & - & $\pm50$ kHz & \cite{Zack2009}\tabularnewline
$14-18$ & $0-4$ & 25 & 25 & 15 & 25 & 5 & $\pm50$ kHz & \cite{Zack2009}\tabularnewline
\hline 
0, 8-9, 13-18 & 0-4 & 34 & 32 & 19 & 30 & 9 & \multicolumn{2}{c}{\textbf{Total \# of data: 124}}\tabularnewline
\end{tabular*}

\vspace{-0.3cm}
\end{table*}

\textcolor{black}{\emph{The experiments. }}The spectra were measured
using a Balle-Flygare type Fourier-Transform microwave spectrometer,
described in detail in \cite{Sun2009}. The instrument consists of
a Fabry-Perot cavity formed from two spherical aluminum mirrors arranged
in a near confocal arrangement. Microwave radiation is injected into
the cavity via an antenna embedded in one mirror, and molecular emission
is detected by an antenna in the opposite mirror. The signals are
recorded in the time domain and are fast Fourier transformed to generate
a frequency-domain spectrum. Each transition appears as a Doppler
doublet and the transition frequencies are reported as the average
of the two Doppler components.

\begin{table*}[t]
\caption{\label{tab:comparisonOfDerivedPhysicalQuantities}\textcolor{black}{\scriptsize Comparison
of some physical quantities derived from our potential, to previous
published values.}}

\begin{tabular*}{1\textwidth}{@{\extracolsep{\fill}}ccccccccc}
\noalign{\vskip1.5mm}
Physical & \multirow{2}{*}{Units} & \multirow{2}{*}{Isotopologue} & This work & Ref. \cite{Zack2009} & Ref. \cite{Moravec2001} & Ref. \cite{Kim2001} & Ref. \cite{Fancher1998} & Ref.\textcolor{black}{{} \cite{Bauschlicher1998} }\tabularnewline
\noalign{\vskip-1.5mm}
\noalign{\vskip1.5mm}
 quantity &  &  & (empirical) & (empirical) & (measured) & (measured) & (measured) & \textcolor{black}{(}\textcolor{black}{\emph{ab initio}}\textcolor{black}{)}\tabularnewline[2mm]
\noalign{\vskip-1.5mm}
\hline 
\noalign{\vskip1mm}
\textcolor{black}{$r_{e}$ } & \textcolor{black}{$\mbox{\AA}$} & $^{64}$Zn$^{16}$O & $1.7046145\pm0.000002$ & $1.7047\pm0.0002$ & $-$ & $-$ & $-$ & $-$\tabularnewline
 &  & $^{66}\mbox{Zn}^{16}\mbox{O}$ & $1.7046151\pm0.000002$ & $1.7047\pm0.0002$ & $-$ & $-$ & $-$ & $-$\tabularnewline
 &  & $^{67}\mbox{Zn}^{16}\mbox{O}$ & $1.7046154\pm0.000002$ & $1.7047\pm0.0002$ & $-$ & $-$ & $-$ & $-$\tabularnewline
 &  & $^{68}\mbox{Zn}^{16}\mbox{O}$ & $1.7046157\pm0.000002$ & $1.7047\pm0.0002$ & $-$ & $-$ & $-$ & $-$\tabularnewline
 &  & $^{70}\mbox{Zn}^{16}\mbox{O}$ & $1.7046162\pm0.000002$ & $1.7047\pm0.0002$ & $-$ & $-$ & $-$ & $-$\tabularnewline
$E_{v=1}-E_{v=0}$ & cm$^{-1}$ & $^{64}$Zn$^{16}$O & $728.395\mbox{\ensuremath{\pm}}0.007$ & $-$ & $726\pm20$  & \textcolor{black}{\footnotesize $770\pm40$ } & $805\pm40$  & \textcolor{black}{\footnotesize $727$}\tabularnewline
\end{tabular*}
\end{table*}

ZnO was synthesized from the reaction of zinc metal vapor and 0.5\%
N$_{2}$O in argon in a discharge assisted laser ablation source \cite{Sun2010}.
The metal vapor was generated by ablating a rotating/translating zinc
rod with the second harmonic (532 nm) of a Nd:YAG laser (100 mJ/5
ns pulse). The gas mixture was introduced into the cavity by a pulsed
valve operating at a 10 Hz rate, entraining the metal vapor before
application of a DC discharge (1.0 kV, 30-50 mA). The molecular beam
was oriented at a 40$^{\circ}$ angle relative to the optical axis,
and the ablation laser was introduced perpendicular to the supersonic
jet. The new transition energy measurements are presented in Table
\ref{tab:newData}.

Adding our new measurements from Table \ref{tab:newData} to the complete
set of high-resolution spectra measured thus far for ZnO, yields the
final dataset presented in Table \ref{tab:fullDataSet}.

\emph{The potential energy function. }The MLR (Morse/Long-range) model
for potential energy functions \cite{LeRoy2009,LeRoy2011} is particularly
pertinent when working with very limited data. In 2011 an analytic
potential was built for a molecular state with an extremely limited
dataset \cite{Dattani2011}, where  it was shown that for the $c(1^{3}\Sigma_{g}^{+})$-state
of Li$_{2}$, the MLR model made it possible to bridge a gap of more
than 150 THz ($>$ 5000 cm$^{-1}$) between data at the very bottom
of the well and data extremely close to the dissociation limit, with
a function that was analytic globally. In 2013, high-resolution measurements
showed that predictions made by that MLR potential, for the energies
in the middle of this gap of more than 5000 cm$^{-1}$, were correct
to about 1 cm$^{-1}$ \cite{Semczuk2013}. A more extreme example
was performed very recently for the $b(1^{3}\Pi_{u})$-state of Li$_{2}$,
for which the region of the potential that had been experimentally
unexplored was roughly the same size as in the case of the $c$-state,
except that rather than bridging a gap between data near the bottom
and data near the top of the well, only data at the bottom of the
well was available \cite{Dattani2014}. Nevertheless, it was still
possible to build  an analytic MLR potential for this state that reproduced
the available data at the very bottom of the well, while also achieving
the correct long-range behavior at the dissociation limit. 

The present case of the $X$-state of ZnO is a much more challenging
case. Two reasons are (1) the extrapolation from the data region to
the dissociation limit is nearly 1 PHz large, and (2) with only pure
rotational data, no vibrational information is available. In this
work we overcame these challenges by calculating accurate long-range
potential terms, and incorporating them into a carefully selected
MLR function. As shown in \cite{LeRoy2009}, all MLR models have the
property: 

\vspace{-0.4cm}

\begin{equation}
\lim_{r\rightarrow\infty}V_{{\rm MLR}}(r)=\mathfrak{D}_{e}-u_{{\rm LR}}(r)+\mathcal{O}\left(u_{{\rm LR}}(r)^{2}\right)+\cdots,\label{eq:V_MLR_longRangeLimit}
\end{equation}
in which $\mathfrak{D}_{e}$ is the dissociation energy and the long-range
function $u_{{\rm LR}}(r)$ is chosen so that Eq. \ref{eq:V_MLR_longRangeLimit}
represents the correct theoretical long-range behavior of the function.
For the $X$-state of ZnO, we choose 

\vspace{-0.4cm}

\begin{equation}
u_{{\rm LR}}=\frac{C_{6}D_{6}(r)}{r^{6}}+\frac{C_{8}D_{8}(r)}{r^{8}}+\frac{C_{10}D_{8}(r)}{r^{10}},
\end{equation}
in which $D_{m}(r)$ are damping functions, as defined in \cite{LeRoy2011}
to take into account electron wavefunction overlap which is not accounted
for by $C_{m}$ constants alone. Upon insertion into Eq. \ref{eq:V_MLR_longRangeLimit}
gives (note $\lim_{r\to\infty}$$D_{m}(r)=1$ for all $m$):

\vspace{-0.4cm}

\begin{equation}
\lim_{r\rightarrow\infty}V_{{\rm MLR}}(r)=\mathfrak{D}_{e}-\frac{C_{6}}{r^{6}}-\frac{C_{8}}{r^{8}}-\frac{C_{10}}{r^{10}}\cdots.
\end{equation}

At present, long-range coefficients for ZnO are only known for bulk
ZnO. Therefore, in this work we calculate $C_{6}$, $C_{8}$ and $C_{10}$
for Zn$(^{1}S)+\mbox{O}(^{1}D)$, which is the unbound state to which
the $X$-state of ZnO dissociates. We use the exchange-hole dipole
moment (XDM) model \cite{becke2007,otero2013}, which is a model
of dispersion based on second-order perturbation theory. The key feature
of XDM is that the source of the instantaneous dipole moments responsible
for the dispersion interaction is taken to be the dipole moment of
the exchange hole. The dispersion coefficients ($C_{m}$) are then
given in terms of the multipole moments of the exchange hole, $\langle M_{\ell}^{2}\rangle$
($\ell=1,2\dots$ for dipole, quadrupole, etc. moments), and the atomic
polarizabilities, $\alpha$. For $C_{6}$, $C_{8}$ and $C_{10}$,
the resulting formulas are given in Eqs. \ref{eq:formulaForC6} to
\ref{eq:formulaForC10} for any atoms $i$ and $j$, in this case
Zn and O. In practice, the Becke-Roussell model of the exchange hole
\cite{becke1989} is used in XDM calculations as it allows straightforward
evaluation of the moments from local properties of the electron density.
The interested reader is directed to Ref. \cite{becke2007} for a
complete description of the model.

\begin{eqnarray}
C_{6}^{ij} & = & \frac{\alpha_{i}\alpha_{j}\langle M_{1}^{2}\rangle_{i}\langle M_{1}^{2}\rangle_{j}}{\langle M_{1}^{2}\rangle_{i}\alpha_{j}+\langle M_{1}^{2}\rangle_{j}\alpha_{i}},\label{eq:formulaForC6}\\
C_{8}^{ij} & = & \frac{3}{2}\frac{\alpha_{i}\alpha_{j}\left(\langle M_{1}^{2}\rangle_{i}\langle M_{2}^{2}\rangle_{j}+\langle M_{2}^{2}\rangle_{i}\langle M_{1}^{2}\rangle_{j}\right)}{\langle M_{1}^{2}\rangle_{i}\alpha_{j}+\langle M_{1}^{2}\rangle_{j}\alpha_{i}},\label{eq:formulaForC8}\\
C_{10}^{ij} & = & 2\frac{\alpha_{i}\alpha_{j}\left(\langle M_{1}^{2}\rangle_{i}\langle M_{3}^{2}\rangle_{j}+\langle M_{3}^{2}\rangle_{i}\langle M_{1}^{2}\rangle_{j}\right)}{\langle M_{1}^{2}\rangle_{i}\alpha_{j}+\langle M_{1}^{2}\rangle_{j}\alpha_{i}}\label{eq:formulaForC10}\\
 &  & +\frac{21}{5}\frac{\alpha_{i}\alpha_{j}\langle M_{2}^{2}\rangle_{i}\langle M_{2}^{2}\rangle_{j}}{\langle M_{1}^{2}\rangle_{i}\alpha_{j}+\langle M_{1}^{2}\rangle_{j}\alpha_{i}}.\nonumber 
\end{eqnarray}

Fully numerical, self-consistent Hartree-Fock calculations were performed
for the Zn and O atoms using the NUMOL program \cite{Becke1990}.
Dispersion coefficients were then calculated from the XDM model using
free atomic polarizabilities obtained from Ref. \cite{crc}. This
approach gives homonuclear $C_{6}$ dispersion coefficients (in atomic
units) of 356.5 for diatomic Zn$(^{1}S$) and 16.14 for diatomic
O$(^{3}P$), in good agreement with the available reference values
of 359 \cite{Qiao2012} and 14.89 \cite{Margoliash1978}, respectively.
Note that the ground state of ZnO will dissociate to give an $\mbox{O}(^{1}D)$
atom and the Zn--O dispersion coefficients were calculated using the
lowest-energy single Slater-determinant reference state for oxygen.
The atomic polarizability of $\mbox{O}(^{1}D)$ was taken to be 5.492
au by correcting the experimental polarizability with the calculated
difference between singlet and triplet results \cite{Medved2000}.
Our final XDM dispersion  coefficients for ZnO, in atomic units,
are $C_{6}=71.34$, $C_{8}=2927$ and $C_{10}=1.368\times10^{5}$. 

\emph{Computational details and results. }We use the diatomic Hamiltonian
model of \cite{LeRoy2009}, with the potential energy function corresponding
to the primary isotopologue $^{64}$Zn$^{16}$O and represented by
the MLR model (described in the previous section of this Letter).
The potentials for all other stable isotopologues are represented
by adding Born-Oppenheimer breakdown (BOB) corrections to the $^{64}$Zn$^{16}$O
potential -- we use the BOB functions of \cite{LeRoy2009}. The final
potentials and BOB correction functions were calculated by a direct-potential-fit
(DPF) to all high-resolution spectroscopic transitions of gas-phase
ZnO measured thus far, as depicted by the dataset summary of Table
\ref{tab:fullDataSet}. The parameters of the potentials and BOB correction
functions were obtained by a least squares fit of the eigenvalues
of the Hamiltonian, to the measured transition energies. These measured
energies are reproduced by our fits to well within their experimental
uncertainties.  All DPF calculations were performed with a freely
available program ${\tt DPotFit}2.0$, and further computational details
are described in the program's user manual \cite{LeRoy2013}. The
parameters defining our final recommended MLR function and BOB correction
function parameters are listed in Table \ref{tab:parametersForPotential}.
Fig. \ref{fig:potential} displays this recommended potential graphically,
and Table \ref{tab:comparisonOfDerivedPhysicalQuantities} compares
physical quantities derived from our potential, to values obtained
in previous papers. 

\emph{Conclusion. }We have demonstrated a technique for building global
empirical potentials using only (very little) pure rotational data,
and our application to gas-phase ZnO has engendered some compelling
results. Without vibrational information, our empirical potential
was able to predict the vibrational energy spacing correctly, and
with more than three orders of magnitude higher precision than the
best experimental measurements and \emph{ab initio} calculations to
date. Our empirical potential which was based on only data spanning
only about $\sim$ 10\% of the well, was in excellent agreement with
\emph{ab initio} calculations over a significantly larger portion
of the experimentally unexplored well. We strongly encourage the use
of this technique for making the most of pure rotational spectra,
and we anticipate it becoming an important tool for microwave spectroscopy. 

\begin{table}
\caption{{\scriptsize Parameters defining our recommended MLR potential for
the $X$-state of $^{64}$Zn$^{16}$O and the BOB correction functions
for $^{64}$Zn$^{16}$O and all the other stable isotopologues. Parameters
in square brackets were held fixed in the fit, while numbers in round
brackets are 95\% confidence limit uncertainties in the last digit(s)
shown. The potential also incorporates damping functions according
to \cite{LeRoy2011}, with $s=-2$ and $\rho=0.88$. \label{tab:parametersForPotential}}}

\begin{tabular*}{1\columnwidth}{@{\extracolsep{\fill}}lc>{\raggedright}p{0.39\columnwidth}lcc}
\textcolor{black}{\footnotesize $\mathfrak{D}_{e}$} & \textcolor{black}{\footnotesize cm$^{-1}$} & \textcolor{black}{\footnotesize $\;[29979.25]$} & \textcolor{black}{\footnotesize $\{p_{{\rm ad}},q_{{\rm ad}}\}$} &  & \textcolor{black}{\footnotesize $\{6,3\}$}\tabularnewline
\textcolor{black}{\footnotesize $r_{e}$} & \textcolor{black}{\footnotesize $\mbox{\AA}$} & \textcolor{black}{\footnotesize $\;1.704682(2)$} & \textcolor{black}{\footnotesize $u_{0}^{{\rm Zn}}$} & \textcolor{black}{\footnotesize cm$^{-1}$} & \textcolor{black}{\footnotesize $-1.2(4)$}\tabularnewline
\textcolor{black}{\footnotesize $C_{6}$} & \textcolor{black}{\footnotesize a.u.} & \textcolor{black}{\footnotesize $\;[71.34]$} & \textcolor{black}{\footnotesize $u_{\infty}^{{\rm Zn}}$} & \textcolor{black}{\footnotesize cm$^{-1}$} & \textcolor{black}{\footnotesize $[0.0]$}\tabularnewline
\textcolor{black}{\footnotesize $C_{8}$} & \textcolor{black}{\footnotesize a.u.} & \textcolor{black}{\footnotesize $\,[2927]$} &  &  & \tabularnewline
\textcolor{black}{\footnotesize $C_{10}$} & \textcolor{black}{\footnotesize a.u.} & \textcolor{black}{\footnotesize $\,[1.368\times10^{5}]$} &  &  & \tabularnewline
\textcolor{black}{\footnotesize $\{p,q\}$} &  & \textcolor{black}{\footnotesize $\,\{25,4\}$} & \textcolor{black}{\footnotesize $\{p_{{\rm na}},q_{{\rm na}}\}$} &  & \textcolor{black}{\footnotesize $\{3,3\}$}\tabularnewline
\textcolor{black}{\footnotesize $r_{{\rm ref}}$} & \textcolor{black}{\footnotesize $\mbox{\AA}$} & \textcolor{black}{\footnotesize $\;[1.75]$} & \textcolor{black}{\footnotesize $t_{0}^{{\rm Zn}}$} &  & \textcolor{black}{\footnotesize $[0.0]$}\tabularnewline
\textcolor{black}{\footnotesize $\beta_{0}$} &  & \textcolor{black}{\footnotesize $0.05598913$} & \textcolor{black}{\footnotesize $t_{1}^{{\rm Zn}}$} &  & \textcolor{black}{\footnotesize $-0.0010$}\tabularnewline
\textcolor{black}{\footnotesize $\beta_{1}$} &  & \textcolor{black}{\footnotesize $-16.182591$} & \textcolor{black}{\footnotesize $t_{\infty}^{{\rm Zn}}$} &  & \textcolor{black}{\footnotesize $[0.0]$}\tabularnewline
\textcolor{black}{\footnotesize $\beta_{2}$} &  & \textcolor{black}{\footnotesize $-100.57307$} &  &  & \tabularnewline
\textcolor{black}{\footnotesize $\beta_{3}$} &  & \textcolor{black}{\footnotesize $-423.085$} &  &  & \tabularnewline
\textcolor{black}{\footnotesize $\beta_{4}$} &  & \textcolor{black}{\footnotesize $-1345.811$} & \textcolor{black}{\footnotesize $t_{0}^{{\rm O}}$} &  & \textcolor{black}{\footnotesize $[0.0]$}\tabularnewline
\textcolor{black}{\footnotesize $\beta_{5}$} &  & \textcolor{black}{\footnotesize $-3509.31$} & \textcolor{black}{\footnotesize $t_{1}^{{\rm O}}$} &  & \textcolor{black}{\footnotesize $0.15$}\tabularnewline
\textcolor{black}{\footnotesize $\beta_{6}$} &  & \textcolor{black}{\footnotesize $-8573.9$} & \textcolor{black}{\footnotesize $t_{2}^{{\rm O}}$} &  & \textcolor{black}{\footnotesize $0.9$}\tabularnewline
\textcolor{black}{\footnotesize $\beta_{7}$} &  & \textcolor{black}{\footnotesize $-20624$} & \textcolor{black}{\footnotesize $t_{3}^{{\rm O}}$} &  & \textcolor{black}{\footnotesize $0.1$}\tabularnewline
\textcolor{black}{\footnotesize $\beta_{8}$} &  & \textcolor{black}{\footnotesize $-36370$} & \textcolor{black}{\footnotesize $t_{\infty}^{{\rm O}}$} &  & \textcolor{black}{\footnotesize $[0.0]$}\tabularnewline
\textcolor{black}{\footnotesize $\beta_{9}$} &  & \textcolor{black}{\footnotesize $-2.87\times10^{4}$} &  &  & \tabularnewline
\end{tabular*}
\end{table}

\textcolor{black}{\emph{Acknowledgments. }}\textcolor{black}{We} would
like to thank Dr. Nick Walker of Newcastle University for informing
us about one of the main selling points of our method. NSD thanks
JSPS and Yoshitaka Tanimura for generous hospitality, RJL thanks NSERC/CRSNG,
and LMZ acknowledges NSF Grant CHE-1057924 for financial support.

\bibliographystyle{apsrev4-1}

\end{document}